\documentclass[aps,longbibliography,showpacs,twocolumn,superscriptaddress,amsmath,amssymb,verbatim]{revtex4-1}

\usepackage[thinlines]{easytable}
\usepackage{graphicx}% Include figure files
\usepackage{lmodern}
\usepackage{epstopdf}
\usepackage{dcolumn}% Align table columns on decimal point
\usepackage{bm}% bold math
\usepackage{subfigure}
\usepackage{makecell}
\usepackage{amsmath} %math
\usepackage[pdftex,colorlinks=true,citecolor=blue,linkcolor=blue,urlcolor=blue,bookmarks=true]{hyperref}
\usepackage{booktabs}
\usepackage{multirow} % Add this package for multirow functionality
\usepackage{array} % For better array manipulation

\usepackage{color}
\usepackage{textcomp}
\definecolor{red}{rgb}{1,0,0}

\definecolor{blue}{rgb}{0,0,1}

\definecolor{green}{rgb}{0,1,0}

\usepackage{multirow}
\usepackage[utf8]{inputenc}
\begin{document}
	\preprint{APS}

\title{Coexistence of static and dynamic local magnetic fields in an $S$ = 3/2 honeycomb lattice antiferromagnet Co$_{2}$Te$_{3}$O$_{8}$
 }

\author{J. Khatua}
\affiliation{Department of Physics, Indian Institute of Technology Madras, Chennai 600036, India}
\affiliation{Department of Physics, Sungkyunkwan University, Suwon 16419, Republic of Korea}
\author{Suheon Lee}
\affiliation{Center for Artificial Low Dimensional Electronic Systems, Institute for Basic Science, Pohang 37673, Republic of Korea}
\author{M. Pregelj }
\affiliation{Jo\v{z}ef Stefan Institute, Jamova c. 39, 1000 Ljubljana, Slovenia}
\affiliation{Faculty of Mathematics and Physics, University of Ljubljana, Jadranska ulica 19, 1000 Ljubljana, Slovenia}
\author{Samiul Sk}
\affiliation{Department of Physics, Bennett University, Greater Noida 201310, Uttar Pradesh, India}
\author{S. K. Panda}
\affiliation{Department of Physics, Bennett University, Greater Noida 201310, Uttar Pradesh, India}
\author{Bassam Hitti}
\affiliation{Centre for Molecular and Materials Science, TRIUMF, Vancouver, British Columbia, Canada V6T 2A3}
\author{Gerald D. Morris}
\affiliation{Centre for Molecular and Materials Science, TRIUMF, Vancouver, British Columbia, Canada V6T 2A3}
\author{ I. da Silva}
\affiliation{ISIS Neutron and Muon Source, Rutherford Appleton Laboratory, Harwell Campus, Didcot, OX11 0QX, United Kingdom}
\author{Kwang-Yong Choi}
\affiliation{Department of Physics, Sungkyunkwan University, Suwon 16419, Republic of Korea}
\author{P. Khuntia}
\email[]{pkhuntia@iitm.ac.in}
\affiliation{Department of Physics, Indian Institute of Technology Madras, Chennai 600036, India}
\affiliation{Quantum Centre of Excellence for Diamond and Emergent Materials, Indian Institute of Technology Madras,
	Chennai 600036, India.}

\date{\today}

\begin{abstract}

	Two-dimensional honeycomb lattices, characterized by their low coordination numbers, provide a fertile platform for exploring various quantum phenomena due to the intricate interplay between competing magnetic interactions, spin-orbit coupling, and crystal electric fields. Beyond the widely studied $J_{\mathrm{eff}}$ = 1/2 honeycomb systems, $S = 3/2$ honeycomb lattices present a promising alternative route to realizing the classical spin liquid-like state within the spin-$S$ Kitaev models. Herein, we present crystal structure, thermodynamic, neutron diffraction and muon spin relaxation ($\mu$SR) measurements, complemented by density functional theory (DFT) calculations on an unexplored $3d$ transition metal based  compound Co$_{2}$Te$_{3}$O$_{8}$, where Co$^{2+}$ ($S = 3/2$) ions form a distorted honeycomb lattice in the crystallographic $bc$-plane without any anti-side disorder between constituent atoms. A clear $\lambda$-type anomaly around $T_{\rm N} \approx 55~\mathrm{K}$ in both magnetic susceptibility and specific heat data indicates the onset of a long-range ordered state below 55 K. The dominant antiferromagnetic interaction between $S$ = 3/2 moments is evidenced by a relatively large negative Curie-Weiss temperature ($\theta_{\mathrm{CW}} = -103~\mathrm{K}$) derived from magnetic susceptibility data and supported by DFT calculations. The signature of long-range antiferomagnetic order state in the thermodynamic data  is corroborated by neutron diffraction and $\mu$SR results. Furthermore, $\mu$SR experiments reveal the coexistence of static and dynamic local magnetic fields below $T_{\rm N}$, along with a complex magnetic structure that can be associated with  \textit{XY}-like antiferromagnet, as confirmed by neutron diffraction experiments.

\end{abstract}
\maketitle
\section{Introduction}
Quantum magnets wherein competing interactions, anisotropic magnetic interactions, crystal electric field, and spin-orbit coupling, are at play provide a suitable venue to realize exotic quantum many-body ground states such as quantum spin liquid (QSL). QSL is an elusive state of matter where electron spins fail to undergo a long-range magnetic order despite strong exchange interactions, characterized by fractionalized excitations and long-range quantum entanglement \cite{Balents2010,KITAEV20062}. To this end, enormous efforts have been devoted towards experimental realization of QSL state with fractionalized excitations such as spinons and Majorana fermions which hold promise for uncovering the mechanisms of high-temperature superconductivity and for practical applications in quantum computing \cite{Savary_2016,ANDERSON1973153,Takagi2019,RevModPhys.80.1083,PhysRevA.79.032311,PhysRevLett.70.4003}. \\  Among the various mechanisms proposed for QSL states in different two-dimensional (2D) lattices, such as triangular, kagome, and honeycomb \cite{KHATUA20231,Khuntia2020,jeon2023oneninth}, the most sought after is the Kitaev QSL in honeycomb lattices, which has generated a flurry of research interest in contemporary condensed matter physics \cite{KITAEV20062}.
Interestingly, the bipartite honeycomb lattice hosts an exactly solvable Kitaev QSL ground state characterized by Majorana fermion and localized $Z_{2}$ fluxes \cite{TREBST20221}. The spin-orbit mediated bond-dependent anisotropic Ising type known as Kitaev interactions between nearest-neighbor spins offer an alternative route to realize frustration induced  QSL state in honeycomb lattices \cite{PhysRevLett.112.077204}. \\ \\
Recently, the quest for Kitaev spin liquid has expanded beyond 4$d$/5$d$ transition metal-based honeycomb lattices \cite{PhysRevLett.108.127204,Do2017,Banerjee2018,doi:10.7566/JPSJ.90.062001,PhysRevB.108.054442} to include several alternative platforms, among which  Co-based (3\textit{d}$^{7}_{\rm free}$, $S$ = 3/2. $L$ = 2) honeycomb lattices have brought  fresh excitement for realizing Kitaev spin liquid  due to their potential to host spin–orbit-entangled $J_{\rm eff}$ = 1/2 moments of Co$^{2+}$ ions \cite{PhysRevLett.125.047201}. This $J_{\rm eff} = 1/2$ pseudospins originates when spin-orbit coupling further splits the ground-state multiplet of the metal–ligand octahedral crystal field, characterized by $S = 3/2$ and an effective orbital moment $L_{\rm eff} = 1$ \cite{PhysRevLett.125.047201}.
 Moreover, when the metal--ligand--metal bond angle approaches $90^\circ$, quantum interference among multiple superexchange pathways between $J_{\rm eff} = 1/2$ pseudospins  can suppress conventional isotropic Heisenberg interactions, favoring the anisotropic Kitaev interaction potential to host excitonic magnetism \cite{adn8694}. Within this framework, recent years have witnessed  the discovery of several cobalt based Kitaev candidates including Na$_{3}$Co$_{2}$SbO$_{6}$ \cite{PhysRevB.107.054411,PhysRevB.109.L060410}, Na$_{2}$Co$_{2}$TeO$_{6}$ \cite{Lin2021} and BaCo$_{2}$(AsO$_{4}$)$_{2}$ \cite{tu2023evidencegaplessquantumspin,Zhang2023}  where the Co$^{2+}$ ions form
edge-sharing  octahedra honeycomb lattices with $J_{\rm eff}$ = 1/2 moments.  However, the ground state of these materials is found to exhibit magnetic order, and the presence of Kitaev interactions in these cobaltates is still a topic of ongoing debate in current research. Interestingly, tuning Hund’s coupling and trigonal crystal field splitting has been proposed as a route to enhance Kitaev interactions, a concept recently implemented in thin films of the honeycomb lattice compound Cu$_{3}$Co$_{2}$SbO$_{6}$  \cite{adn8694,PhysRevLett.125.047201}. Beyond the search for QSL,  cobaltates based on $J_{\rm eff} = 1/2$ moments are highly sensitive to external perturbations such as magnetic fields or chemical pressure, often giving rise to intriguing many-body quantum phenomena. Noteworthy examples include field-induced quantum criticality in CoNb$_2$O$_6$ \cite{Liang2015}, quantized magnetization plateaus in triangular lattice compounds Ba$_3$CoSb$_2$O$_9$ \cite{PhysRevLett.110.267201} and Ba$_2$CoTeO$_6$ \cite{PhysRevB.93.094420}, a quantum disordered ground state in BaCo$_2$(P$_{1-x}$V$_x$)$_2$O$_8$ \cite{PhysRevB.98.220407}, and a field-induced supersolid phase in Na$_2$BaCo(PO$_4$)$_2$ \cite{Gao2022}.\par
While bond-dependent Kitaev interactions in honeycomb lattices offer a platform for exotic fermionic excitations such as Majorana fermions, \textit{XY} antiferromagnets on honeycomb lattices have attracted significant attention for hosting emergent bosonic spin excitations, such as topological magnons \cite{PhysRevX.8.011010,Bao2018}. These bosonic modes arise from magnon band structures, analogous to the formation of topological insulators from electronic band structures \cite{rameshti2022cavity}. For example, in the honeycomb lattice material CoTiO$_3$, inelastic neutron scattering experiments have identified magnon Dirac cones associated with \textit{XY} antiferromagnetic order \cite{PhysRevX.10.011062}. Moreover, \textit{XY} honeycomb lattices serve as spin analogues of superfluid helium and superconductors, supporting Goldstone excitations in the long-wavelength limit \cite{RevModPhys.85.219,auerbach2012interacting}. Experimental realizations of $S = 3/2$ honeycomb lattices are rare; however, weak spin-orbit coupling in cobaltates can stabilize an $S = 3/2$ ground state for Co$^{2+}$ ions. In recent years, apart from the $J_{\rm eff}$ = 1/2 honeycomb lattices, the search for a classical version of the Kitaev spin liquid state has extended to spin $S$ $>$ 1/2 honeycomb lattices, which can also host $Z_{2}$ flux \cite{PhysRevB.98.214404,PhysRevE.82.031113,PhysRevLett.98.247201,PhysRevLett.124.087205,PhysRevB.108.075111}. Interestingly, a possible scenario of  a chiral Kitaev spin liquid state has been proposed in $S$ = 3/2 honeycomb lattice with conserved $Z_{2}$ fluxes for each elementary hexagon \cite{Jin2022}. \\ \\
To identify suitable candidates for realizing classical Kitaev models, it is crucial to explore promising $S = 3/2$ representatives. Herein, we report thermodynamic and microscopic results, corroborated by DFT calculations, on the previously unexplored cobaltate Co$_{2}$Te$_{3}$O$_{8}$ (hereafter CTO), where Co$^{2+}$ ions form a two-dimensional honeycomb lattice stacked along the $a$-axis. A $\lambda$-type anomaly in magnetic susceptibility and specific heat indicates the onset of long-range magnetic order around $T_{\rm N} = 55$~K, consistent with neutron diffraction and $\mu$SR measurements. $\mu$SR results further reveal the coexistence of static and dynamic local magnetic fields below $T_{\rm N}$, characterized by persistent spin dynamics and an order-parameter-like evolution of the local fields, indicative of  \textit{XY}-like antiferromagnet  that is corroborated by neutron diffraction experiments. DFT calculations confirm that long-range magnetic order is stabilized by further-neighbor intra-planar and weak inter-layer antiferromagnetic interactions, underlining the role of moderate spin frustration.

\section{EXPERIMENTAL AND THEORETICAL DETAILS}
Polycrystalline samples of CTO were prepared by a
conventional solid-state reaction of CoCl$_{2}$·
$6$H$_{2}$O  (Alfa Aesar, 99.997 \text{\%}), and
TeO$_{2}$ (Alfa Aesar, 99.97 \text{\%}). The reactants were weighed at an appropriate molar ratio and mixed thoroughly to achieve a homogeneous mixture. Subsequently, the stoichiometric mixture was pressed into pellets, placed in an aluminum crucible, and heated at 400$^\circ$C for 20 hrs. 
This sintering process was repeated at several intermediate temperatures before the final annealing step at 800$^{\circ}$C for 48 hrs
to ensure a single-phase composition. Powder X-ray diffraction (XRD) data were collected at room temperature using a SmartLab Rigaku diffractometer with Cu-$\alpha$ radiation ($\lambda$ = 1.54 Å).\\ \\ Magnetic measurements were performed  using a superconducting quantum interference device vibrating-sample magnetometer (SQUID-VSM, Quantum Design, USA) in the temperature range 2 K $\leq$ $T$ $\leq$ 300 K and in magnetic fields up to 7 T. Specific heat measurements were carried out using a standard relaxation method with a physical property measurement system (PPMS, Quantum Design, USA) in the temperature range 2 K $\leq$ $T$ $\leq$ 300 K and in two magnetic fields, 0 T and 9 T. \\ \\
$\mu$SR measurements were conducted using the surface muon beamline (M20) at the Centre for Molecular and Material Science, TRIUMF, in the temperature range 1.9 K $\leq$ $T$ $\leq$ 90 K in zero-field. During the M20 measurements, the LAMPF spectrometer, coupled with a $^4$He flow cryostat, was utilized to achieve the base temperature of 1.9 K. Approximately 0.5 g of powder samples were placed into a thin Mylar tape envelope, coated with aluminum (about 50 $\mu$m thick), to minimize background signals, and mounted on a copper fork sample stick. The $\mu$SR data collected were analyzed using the musrfit software package \cite{SUTER201269}. Neutron powder diffraction data were collected on the GEM time-of-flight diffractometer at the ISIS Pulsed Neutron and Muon Source of the Rutherford Appleton Laboratory \cite{HANNON200588,ISISN}.
\\ \\
We performed first-principles calculations based on density functional theory (DFT) to investigate the electronic structure and magnetism of CTO, aiming to provide theoretical insights into the experimentally observed magnetic behavior. The total energy corresponding to different magnetic states and the electronic density of states were computed using projector augmented wave (PAW) method~\cite{PhysRevB.50.17953}, as implemented in the Vienna ab initio simulation package (VASP)~\cite{PhysRevB.47.558,PhysRevB.54.11169}. For the exchange and correlation functional, we employed generalized gradient approximation (GGA) of Perdew-Burke-Ernzerhof (PBE)~\cite{GGA}, supplemented by an on-site Hubbard $U$ correction (GGA+U approach)~\cite{GGA+U} to appropriately account for the strong electronic correlations associated with the Co 3$d$ orbitals. A Hubbard $U$ = 6.0 eV and Hund's exchange $J$ = 0.8 eV are considered for Co. A similar value has been reported for Co-d orbitals in CoO based on cRPA calculations \cite{PhysRevB.96.045137}.  The plane-wave energy cutoff of 550 eV was applied, and the Brillouin zone was sampled using a $\Gamma$-centered Monkhorst-Pack mesh of 4$\times$ 8 $\times$ 4. 
\\ \\
To obtain a quantitative description of the magnetic exchange network that underpins the experimentally observed magnetic ground state, we further employed the full-potential linear muffin-tin orbital (FP-LMTO) method~\cite{PhysRevB.12.3060,PhysRevB.36.3809}, as implemented in the RSPt code~\cite{10.1007/3-540-46437-9_4}. From the self-consistent GGA+U electronic structure, the interatomic exchange interactions ($J_{ij}$) between Co spins at $i^{th}$ and $j^{th}$ site were extracted using the magnetic force theorem~\cite{PhysRevB.91.125133}. In this method, the total converged energies of the magnetic system were mapped onto the following Heisenberg spin Hamiltonian
\begin{equation}
\ H =  -\sum_{i\neq j}J_{ij}(\hat{S}_i \cdot \hat{S}_j), 
\end{equation}
where $\hat {S}_i$ \text{and} $\hat {S}_j$ are spin unit vectors for the ${i}^{th}$ and ${j}^{th}$ site, respectively. Finally, the effective $J_{ij}$s are extracted in a linear-response manner via a Green’s function technique using the following formula: 
\begin{equation}
J_{i j}=\frac{\mathrm{T}}{4} \sum_n \operatorname{Tr}\left[\hat{\Delta}_i\left(i \omega_n\right) \hat{G}_{i j}^{\uparrow}\left(i \omega_n\right) \hat{\Delta}_j\left(i \omega_n\right) \hat{G}_{j i}^{\downarrow}\left(i \omega_n\right)\right],
\end{equation}
where $\Delta_i$ is the onsite spin splitting, $G_{ij}$ is the spin-dependent intersite Green’s function, and $\omega_n$ are the $n^{th}$ Fermionic Matsubara frequencies. A detailed discussion of the implementation of the magnetic force theorem in RSPt is provided in Ref.~\onlinecite{PhysRevB.91.125133}. This method has been successfully used for many other transition metal compounds~\cite{PhysRevB.109.035125, PhysRevB.94.064427, PhysRevB.106.L180408}.
 \begin{figure*}
	\centering
	\hspace{0cm}
	\includegraphics[width=0.8\textwidth]{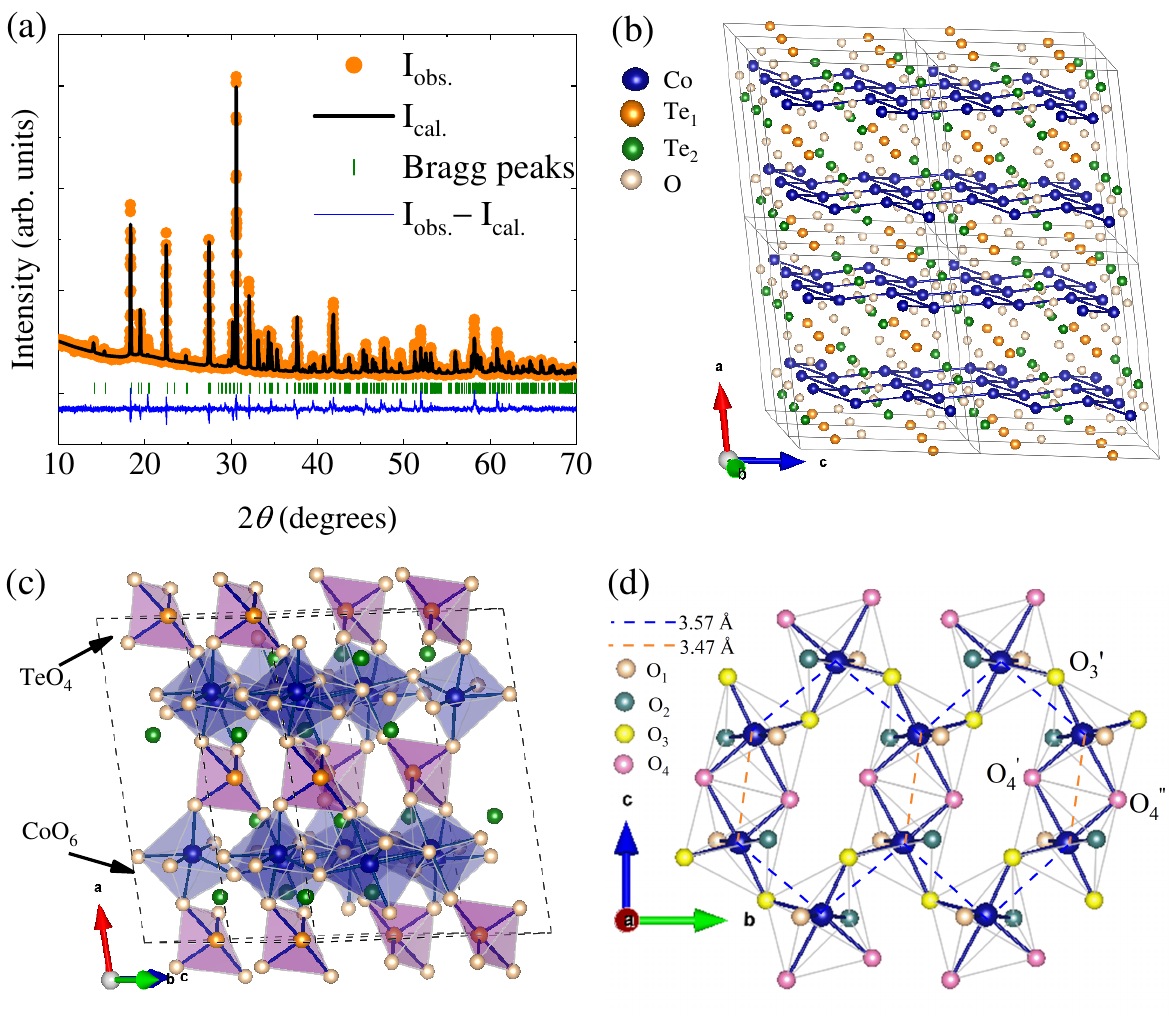}
	\caption{(a) Rietveld refinement profile of powder X-ray diffraction data at room temperature. The solid black line represents the calculated intensity (I$_{\rm cal.}$) for the monoclinic crystal structure of Co$_{2}$Te$_{3}$O$_{8}$, fitting the observed experimental points (I$_{\rm obs.}$). Olive vertical bars indicate the positions of the Bragg reflections, while the difference between calculated and observed intensity  is shown by the blue solid line. (b) Four-unit cells of Co$_{2}$Te$_{3}$O$_{8}$, where Co$^{2+}$ ions form a distorted honeycomb lattice perpendicular to the crystallographic $a$-axis. (c) Schematic shows the nearest-neighbor O$^{2-}$ ions of Co$^{2+}$ form a layer of CoO$_{6}$ distorted octahedra sandwiched between the layer of TeO$_{4}$ distorted square planar structure. (d) A single distorted honeycomb plane of Co$^{2+}$ ions perpendicular to the crystallographic $a$-axis that is composed of corner-sharing Co$_{2}$O$_{10}$  dimer.  }{\label{CTOFig1}}.
\end{figure*}
\section{Results and discussion}
\subsection{Rietveld refinement and crystal structure}
In order to confirm the phase purity and crystal structure, Rietveld refinements of powder X-ray diffraction data were performed using the GSAS software. Figure \ref{CTOFig1}(a) shows the profile of Rietveld refinement that suggests the absence of any detectable secondary phase in the polycrystalline samples of CTO and it crystallizes in the monoclinic crystal structure with space group $C2/c$ \cite{FEGER1999246}. The obtained structural parameters and goodness factors are tabulated in table \ref{table}.  Rietveld refinement reveals the absence of any anti-site disorder between constituent atoms.  Shown in Fig.~\ref{CTOFig1}(b) is the crystal structure of CTO where Co$^{2+}$ ions form a distorted honeycomb lattice in the $bc$-plane. 
In the $bc$-plane, Co$^{2+}$ ions, surrounded by oxygen ions, form a layer of distorted CoO${_6}$ octahedra sandwiched by TeO${_4}$ distorted square planar units (Fig.~\ref{CTOFig1}(c)). To gain deeper insight into the pathways of magnetic exchange interactions among $S$ = 3/2 moments of Co$^{2+}$ ions, Figure~\ref{CTOFig1}(d) presents a magnified view of a single layer of a distorted honeycomb lattice surrounded by oxygen ions. 
 \begin{table}
  	\caption{\label{table}  The structural parameters of CTO were obtained through Rietveld refinement analysis of X-ray diffraction data collected at 300 K. (Space group: $C2/c$, $a$ = 12.675 \AA , $b$ = 5.208 \AA, $c$ = 11.63 \AA, $\alpha = \gamma = 90^{\circ}$, $ \beta$ = 99.01 $^\circ$
  		and $\chi^{2}$ = 2.277, R$_{\rm wp}$ = 6.24 \text{\%}, R$_{\rm p}$ = 4.57 \text{\%}, and R$ _{\rm exp}$ = 4.13\text{\%})}
  	\begin{tabular}{c c c c c  c c} % <-- Alignments: l for left, c for center, and r for right, with vertical lines in between
  		\hline \hline
  		Atom & Wyckoff position & \textit{x} & \textit{y} &\textit{ z}& Occ.\\
  		\hline 
  		Co & 8$f$ & 0.269 \ \ & 0.296 \ \ & 0.150 \ \ & 1 \\
  		Te$_{1}$ & 8$f$ &  0.5 & 0.649 & 0.248  \ \ & 1 \\
  	Te$_{2}$ & 8$f$ & 0.362 & 0.298 & 0.442 & 1\\
  		O$_{1}$& 8$f$ & 0.420 & 0.428 & 0.144 & 1 \\
  		O$_{2}$ & 8$f$ & 0.385 & 0.624 & 0.359 & 1\\
  		O$_{3}$ & 8$f$ & 0.297 & 0.133 & 0.309 & 1\\
  		O$_{4}$ & 8$f$&  0.237 & 0.489 & 0.470 & 1 \\	
  		\hline
  	\end{tabular}
  	%\end{ruledtabular}
  \end{table}
\begin{table}
	\caption{\label{tableangle}  Bond lengths and angles between atoms that lead to dominant antiferromagnetic interactions between \textit{S} = 3/2 moments of Co$^{2+}$ ions in Co$_{2}$Te$_{3}$O$_{8}$ (See Fig.~\ref{CTOFig1}(d)). }
	\begin{tabular}{c c c c c  c c}
		\hline \hline
		Bond length (\AA)  & Bond angle ($^\circ$) &\\
		\hline 
		Co-O$_{4}^{'}$ =  2.115, O$_{4}^{'}$-Co = 2.352  & $\angle$ Co-O$_{4}^{'}$-Co = 102.16 & \\
		Co-O$_{4}^{''}$ =  2.352, O$_{4}^{''}$-Co = 2.115  & $\angle$ Co-O$_{4}^{''}$-Co = 102.16 & \\ 
		Co-O$_{3}^{'}$ =  2.032, O$_{3}^{'}$-Co = 2.015  & $\angle$ Co-O$_{3}^{'}$-Co = 123.998 & \\ 
		\\	
		\hline
	\end{tabular}
	%\end{ruledtabular}
\end{table} 
 \begin{figure*}
	\centering
	\includegraphics[width=\textwidth]{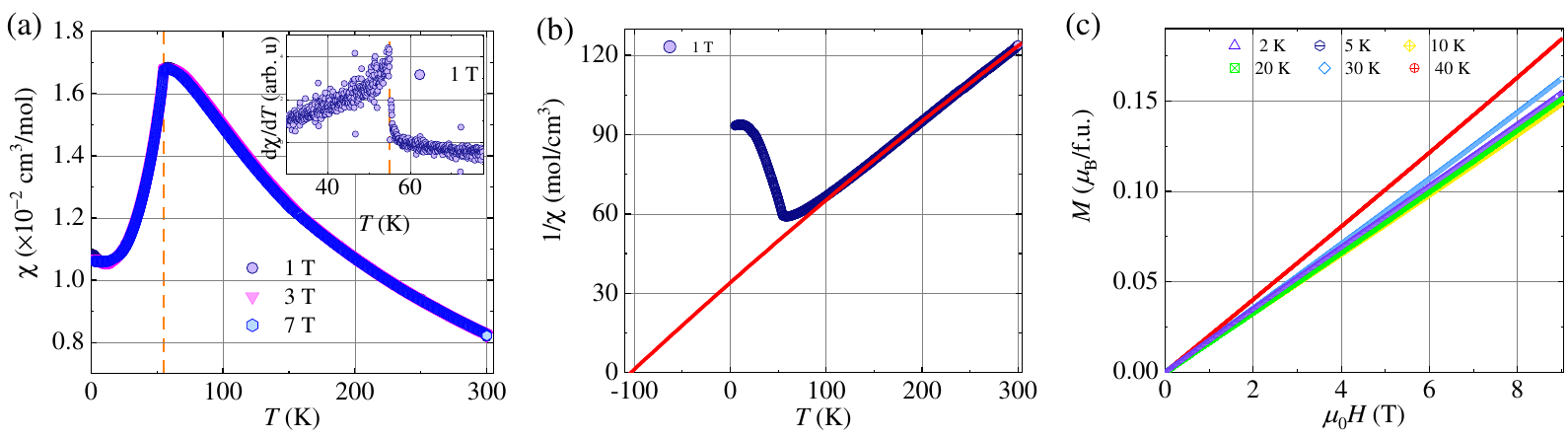}
	\caption{(a) Temperature dependence of magnetic susceptibility in several magnetic fields.  Inset shows the derivative of magnetic susceptibility as a function of temperature in a magnetic field $\mu_{0}H$ = 1 T. The dashed vertical lines mark the temperature at which antiferromagnetic phase transition occurs. (b) Temperature dependence of inverse magnetic susceptibility in a magnetic field $\mu_{0}H$ = 1 T. The solid red line depicts the Curie-Weiss fit. (c) Isotherm magnetization as a function of magnetic field at several temperatures.  }{\label{CTOFig3}}.
\end{figure*}
In the crystal structure of CTO, the nearest-neighbor (3.47 \AA) Co$^{2+}$ ions with surrounded oxygen ions form Co$_{2}$O$_{10}$ dimer. This dimer consists of two edge shared CoO$_{6}$ octahedra running along the crystallographic $c$-axis. The intra-dimer (orange dashed line; Fig.~\ref{CTOFig1}(d)) magnetic exchange interactions between $S$ = 3/2 moments are expected through the superexchange path Co-O$_{4}$-Co ($\angle$102.16$^{\circ}$). The bond length and angles associated with the Co-O-Co superexchanges pathway are listed in Table~\ref{tableangle}. In the unit cell, the Co$_{2}$O$_{10}$ dimers  are connected (blue dashed line; Fig.~\ref{CTOFig1}(d)) through the common oxygen ions and forms inter-dimer exchange path Co-O$_{3}$-Co ($\angle$123.998$^{\circ}$). This combinations of intra-dimer and inter-dimer exchange interactions might lead to  the overall honeycomb-like magnetic lattice of CTO. Despite the slight difference between the intra-dimer and inter-dimer bond lengths, a more pronounced difference in the angles formed by the metal-ligand-metal bonds (Co-O-Co) can result in frustrated exchange interactions  in this bipartite spin-lattice. 
\subsection{Magnetic susceptibility}
Figure \ref{CTOFig3}(a) shows the temperature dependence of magnetic susceptibility, $\chi(T)$, down to 2 K in several magnetic fields. Upon lowering the temperature, $\chi(T)$ gradually increases until  a broad hump just above 55 K (dashed vertical line in Fig.~\ref{CTOFig3}(a)). This indicates the presence of short-range spin correlations typical of low-dimensional magnets, just before the onset of long-range magnetic ordered state. Further lowering the temperature below 55 K, $\chi(T)$ begins to decrease, suggesting the development of antiferromagnetic long-range order. In order to determine the transition temperature from the paramagnetic to the long-range magnetic ordered state, the temperature derivative of $\chi(T)$ was calculated, as shown in the inset of Fig.~\ref{CTOFig3}(a) as a function of temperature. The presence of a clear $\lambda$-type anomaly in $d\chi(T)/dT$ at $T_{\text{N}} \approx 55$ K corresponds to the transition temperature in the present compound, which is consistent with the anomaly in the specific heat data discussed in the following section.\\ \\
The temperature dependence of inverse magnetic susceptibility, 1/$\chi(T)$, is shown in Fig.~\ref{CTOFig3}(b). 
In order to get an idea about  the presence of the dominant magnetic interactions between Co$^{2+}$ moments, the linear region (150 K $\leq$ $T$ $\leq$ 300 K) of 1/$\chi(T)$ data were  fitted by the Curie-Weiss (CW) law, $\chi(T)$ = $\chi_{0}$ + $C$ /($T$ $-$ $\theta_{\rm CW}$). Here, $\chi_{0}$ accounts for the temperature-independent contributions due to core diamagnetism and Van Vleck paramagnetism, $C$ is the Curie constant, and $\theta_{\rm CW}$ is the CW temperature representing the characteristic exchange interaction between Co$^{2+}$ magnetic moments in the spin-lattice. The solid red line in Fig.~\ref{CTOFig3}(b) is the corresponding CW fit which yields $\chi_{0}$ = 7.314 $\times$ 10$^{-4}$ cm$^{3}$/mol, $C$ = 2.97 $\pm$ 0.01 cm$^{3}$ K/mol, and $\theta_{\rm CW}$ = $-$ 103 $\pm$ 1 K.  After subtracting the core diamagnetic contribution $\chi_{\rm dia} = -3.6 \times 10^{-6}$ cm$^3$/mol \cite{Bain2008}, the Van Vleck susceptibility is estimated as $\chi_{\rm VV} = (\chi_{0} - \chi_{\rm dia}) = 7.6 \times 10^{-4}$ cm$^3$/mol. The value of $C$ leads to an effective magnetic moment, $\mu_{\rm eff}$ = $\sqrt{8C}$ = 4.90 $\mu_{\rm B}$, which is little higher than the the expected $\mu_{\rm eff}$ = 3.87 $\mu_{\rm B}$ for free Co$^{2+}$ ions (3$d^{7}$, $S$ = 3/2, L = 2) with high-spin $S$ = 3/2 \cite{PhysRevB.96.064413}. The excess $\mu_{\rm eff}$ ($\approx$ 1 $\mu_{\rm B}$) can be assigned to the presence of partial unquenched orbital moment similar to that observed in other cobaltates \cite{PhysRevB.102.224411,PhysRevB.103.214447}. The  obtained large and negative CW temperature suggests the presence of dominant antiferromagnetic interactions  between $S$ = 3/2 moments of Co$^{2+}$ ions. Moreover, it is noted that below 130 K, the Curie-Weiss fit begins to diverge from the measured 1/$\chi$($T$) data, suggesting that antiferromagnetic spin correlations begin to develop well before reaching the transition temperature $T_{\rm N}$. \\ \\
To further confirm the presence of dominant antiferromagnetic interactions, isotherm magnetization measurements were performed as a function of magnetic field at several temperatures. At 2\,K, five-quadrant magnetization measurements confirmed the absence of hysteresis associated with ferromagnetic contributions. Consequently, only the linear field dependence of magnetization in the first quadrant is shown in
  Fig.~\ref{CTOFig3}(c).  The temperature dependence of isothermal magnetization below $T_{\rm N}$ is consistent with the temperature dependency of $\chi{(T)}$ (see Fig.~\ref{CTOFig3}(a)).
  \begin{figure*}
  	\centering
  	\includegraphics[width=\textwidth]{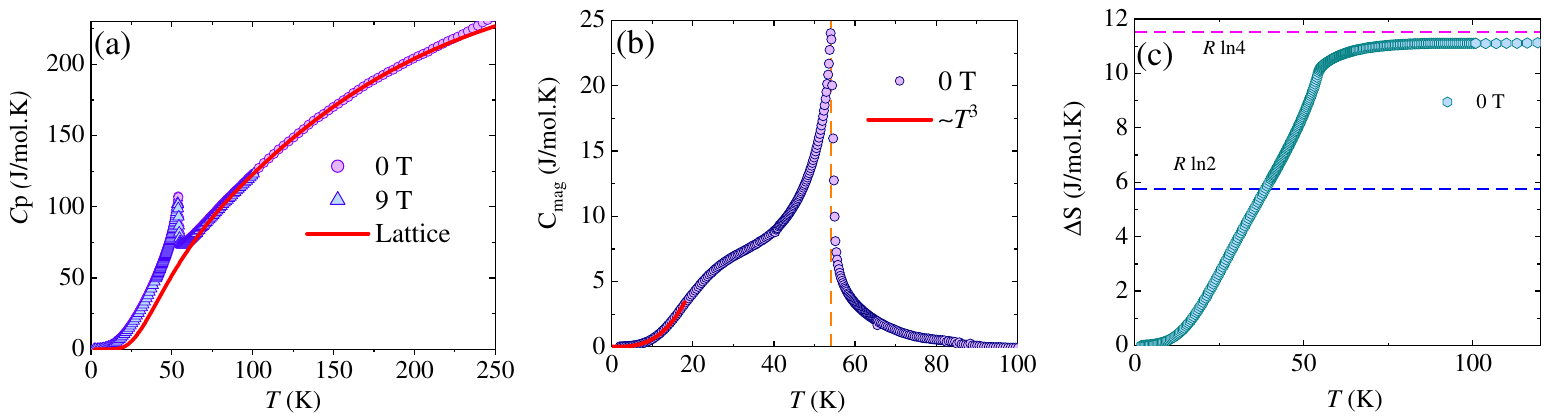}
  	\caption{Temperature dependence of specific heat in two magnetic fields where the solid line accounts for the lattice contributions due to phonons. (b) Temperature dependence of magnetic specific heat with an anomaly at $T_{\rm N}$ indicated by dashed vertical line. Well below $T_{\rm N}$, the magnon excitations are described by a $T^{3}$
  		behavior, as indicated by the solid line. (c) Temperature dependence of magnetic entropy change in a zero magnetic field. The pink and blue horizontal lines show the expected entropy release for high-spin ($S$ = 3/2) and low-spin ($J_{\rm eff}$ = 1/2) Co$^{2+}$ ions, respectively. }{\label{CTOFig4}}.
  \end{figure*}  
\subsection{Specific heat}
Specific heat is an excellent tool to track the magnetic phase transition and the low-energy magnetic excitations in strongly correlated systems. Figure \ref{CTOFig4}(a) depicts the temperature dependence of specific heat, $C_{\rm p}(T)$, in two magnetic fields down to 2 K. A $\lambda$-like anomaly at $T_{\rm N}$ = 55 K is attributed to the presence of long-range magnetic order in CTO. The measured $C_{\rm p}(T)$  can be assigned to the sum of magnetic specific heat ($C_{\rm mag}$) due to spin of Co$^{2+}$ ions and  lattice specific heat ($C_{\rm latt}$) due to phonons. In order to subtract the phonon contribution, the temperature dependence of $C_{\rm p}(T)$ was fitted at high temperatures using a combination of one Debye and two Einstein terms, given by  
\begin{align}
C_{\rm latt} &= C_{D} \left[ 9R \left( \frac{T}{\theta_{D}} \right)^{3} 
\int_{0}^{\theta_{D}/T} \frac{x^{4} e^{x}}{(e^{x} - 1)^{2}} \, dx \right] \nonumber \\
&\quad + \sum_{i = 1}^{2}C_{E_{i}} \left[3 R \left( \frac{\theta_{E_{i}}}{T} \right)^{2} 
\frac{e^{\theta_{E_{i}}/T}}{(e^{\theta_{E_i}/T} - 1)^{2}} \right],
\end{align}
where $\theta_{D}$ is the Debye temperature, $\theta_{E_{i}}$ represents the Einstein temperatures, $R$ is the molar
gas constant. The best fit (red line in Fig.~\ref{CTOFig4} (a)) in the temperature range 120 K $\leq$  $T$ $\leq$ 200 K was achieved using the parameters $\theta_{D}$ = 230 K $\pm$ 5 K, $\theta_{E_{1}}$ = 210 K $\pm$ 4 K, and $\theta_{E_{2}}$ = 640 $\pm$ 10 K  and this fit was then extrapolated down to 2 K. The fitted coefficients $C_{D}$ and $C_{E_{i}}$ were approximately 3, 6, and 4 and their sum is equal to the total number of atoms in CTO \cite{PhysRevB.110.184402}.\\ \\
\begin{figure*}
	\centering
	\includegraphics[width=\textwidth]{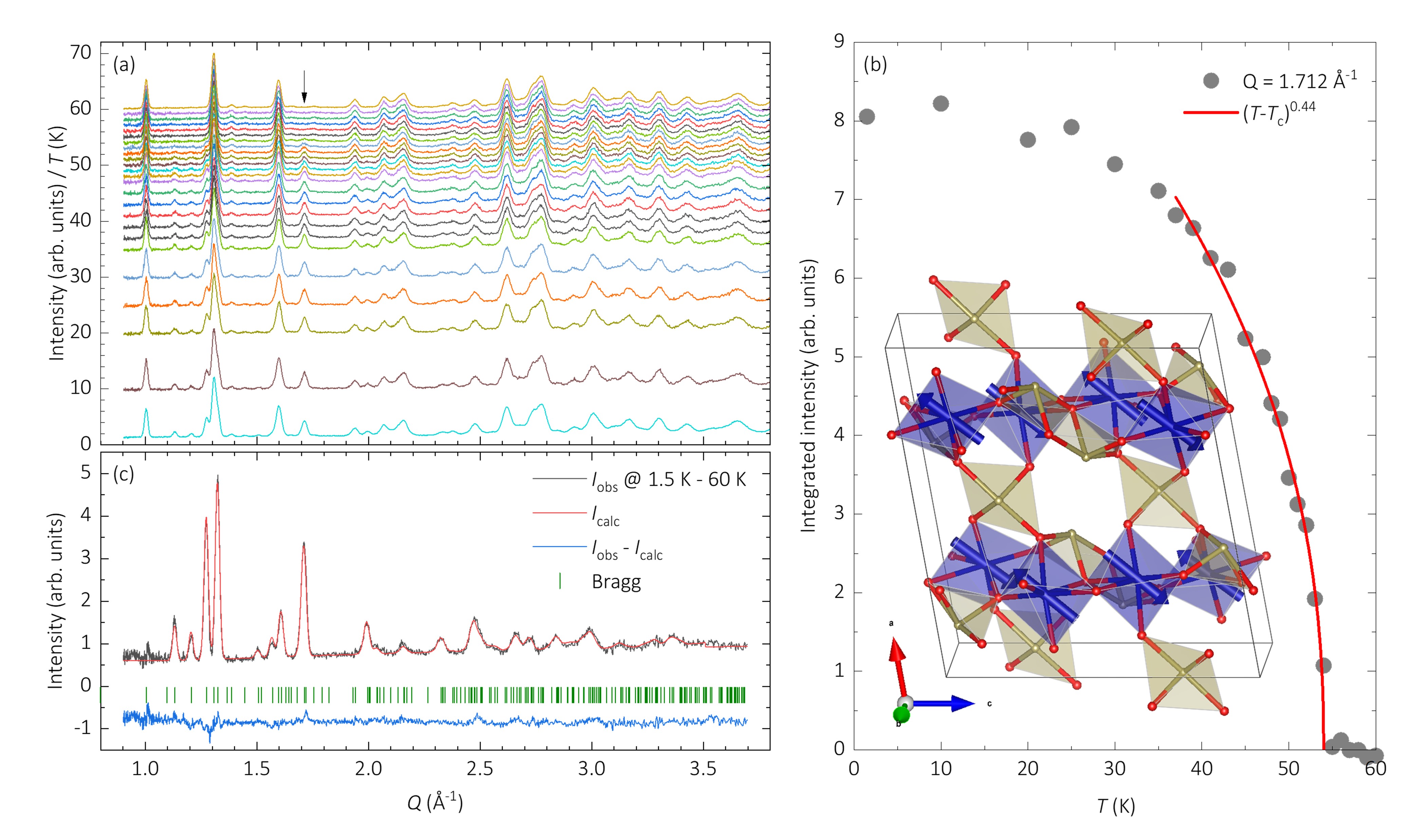}
	\caption{(a) Neutron powder diffraction patterns for Co$_2$Te$_3$O$_8$ compound measured between 1.5 and 60\,K. The intensity scale also correspond to the temperature of the measured diffraction patterns. The arrow denotes the position of the most pronounced magnetic reflections. (b) The derived temperature dependence of the marked-magnetic-peak intensity. The inset shows the refined magnetic structure. (c) The difference between the data measured at 1.5\,K and 60\,K compared with the calculated diffraction pattern and corresponding residual.}
	\label{fig-all}
\end{figure*}
\begin{table*}
	\centering
	\caption{The irreducible representations (IRR) $\Gamma_i$, $i$\,=\,1-4 and corresponding basis vectors $\psi_i^j$ in the $abc$ coordinate system for the space group $C2/c$ appearing in the magnetic representation with $\mathbf{k}=(1,\:0,\:0)$ for the magnetic site (0.269,~0.296,~0.15) of Co atom\,1 and its three crystallographically equivalent sites: $(-0.269,\,0.296,\,0.35)$ of atom\,2, $(-0.269,\,-0.296,\,-0.15)$ of atom\,3, and $(0.269,\,-0.296,\,0.65)$ of atom\,4, with magnetic unit cell corresponding to the $C$\,$-1$ space group. We note that representation analysis was performed using BasIreps program incorporated in the FullProf Suite \cite{rodriguez1990fullprof}.}
	\begin{tabular*}{\linewidth}{@{\extracolsep{\fill}} c   c   c c c  c  c c c  c  c c c  c  c c c }
		\hline \hline
		IRR & Basis & \multicolumn{3}{c}{Atom 1} & & \multicolumn{3}{c}{Atom 2} & & \multicolumn{3}{c}{Atom 3} & & \multicolumn{3}{c}{Atom 4} \\
		& vector & $m_a$ & $m_b$ & $m_c$ & & $m_a$ & $m_b$ & $m_c$ & & $m_a$ & $m_b$ & $m_c$ & & $m_a$ & $m_b$ & $m_c$  \\
		\hline      
		$\Gamma_1$ & $\psi_1^1$ &  1  &  0  &  0  &   & $-1$&  0  &  0  &   &  1  &  0  &  0  &   & $-1$&  0  &  0   \\
		& $\psi_1^2$ &  0  &  1  &  0  &   &  0  &  1  &  0  &   &  0  &  1  &  0  &   &  0  &  1  &  0   \\
		& $\psi_1^3$ &  0  &  0  &  1  &   &  0  &  0  & $-1$&   &  0  &  0  &  1  &   &  0  &  0  & $-1$ \\
		\hline      
		$\Gamma_2$ & $\psi_2^1$ &  1  &  0  &  0  &   & $-1$&  0  &  0  &   & $-1$&  0  &  0  &   &  1  &  0  &  0   \\
		& $\psi_2^2$ &  0  &  1  &  0  &   &  0  &  1  &  0  &   &  0  & $-1$&  0  &   &  0  & $-1$&  0   \\
		& $\psi_2^3$ &  0  &  0  &  1  &   &  0  &  0  & $-1$&   &  0  &  0  & $-1$&   &  0  &  0  &  1   \\
		\hline      
		$\Gamma_3$ & $\psi_3^1$ &  1  &  0  &  0  &   &  1  &  0  &  0  &   &  1  &  0  &  0  &   &  1  &  0  &  0   \\
		& $\psi_3^2$ &  0  &  1  &  0  &   &  0  & $-1$&  0  &   &  0  &  1  &  0  &   &  0  & $-1$&  0   \\
		& $\psi_3^3$ &  0  &  0  &  1  &   &  0  &  0  &  1  &   &  0  &  0  &  1  &   &  0  &  0  &  1   \\
		\hline      
		$\Gamma_4$ & $\psi_4^1$ &  1  &  0  &  0  &   &  1  &  0  &  0  &   & $-1$&  0  &  0  &   & $-1$&  0  &  0   \\
		& $\psi_4^2$ &  0  &  1  &  0  &   &  0  & $-1$&  0  &   &  0  & $-1$&  0  &   &  0  &  1  &  0   \\
		& $\psi_4^3$ &  0  &  0  &  1  &   &  0  &  0  &  1  &   &  0  &  0  & $-1$&   &  0  &  0  & $-1$ \\	   
		\hline \hline
	\end{tabular*}
	\label{tab-irreps}
\end{table*} 
The $C_{\rm mag}(T)$ was determined after subtracting the lattice contributions as shown in Fig.~\ref{CTOFig4}(b) as a function of temperature, which demonstrates a clear $\lambda$-like anomaly $T_{\rm N}$ consistent with that observed anomaly in $d\chi(T)$/$dT$ data (see Fig.~\ref{CTOFig3}(a)). The increase in $C_{\rm mag}$ 
upon lowering the temperature below 100 K ($\approx$ $|\theta_{\rm CW}|$) indicates a slowing down  of spin fluctuations associated with the onset of magnetic ordering. Interestingly, below $T<<T_{N}$, $C_{\rm mag}(T)$ follows $\alpha T^{3}$ behavior behavior (red line in Fig.~\ref{CTOFig4}(b)), indicating the presence of magnon excitations that is typical of long-range ordered magnets. To further confirm the high-spin moment of Co$^{2+}$ ions and quantify the percentage of entropy release at $T_{\rm N}$, the temperature dependence of entropy change ($\Delta S$) was calculated by integrating $C_{\rm mag}(T)/T$ over the temperature range from 2 K to 150 K, as shown in Fig.~\ref{CTOFig4}(c). We observed that the estimated magnetic entropy saturates to the value  11.52 J/mol.K that is close to the expected entropy ($R$ln(2$S$+1)) for the \textit{S} = 3/2 spin of Co$^{2+}$ ions.  Furthermore, at $T_{\rm N}$, the 89 \text{\%} of the saturation
value  of entropy is  released, indicating the presence of moderate spin frustration. On the other hand, the remaining 11\text{\%} of the entropy released well above $T_{\rm N}$ may be attributed to short-range spin correlations or a possible overestimation of the lattice contribution given the simplicity of the model considered here, due to the non- availability of a magnetic analog in extracting lattice specific heat.
\subsection{Neutron-powder diffraction}
To shed light on the magnetic ordered state and the associated order parameter in this distorted honeycomb lattice antiferromagnet, neutron powder diffraction experiments were conducted.
The  diffraction patterns were measured between 1.5 and 60\,K and are shown in Fig.~\ref{fig-all}(a).
The diffraction pattern measured at 60\,K can be very well refined to the expected crystal structure, yielding R$_{\rm f}$-factor of 3.95.
Below $\sim$53\,K additional reflections emerge (arrow in Fig.\,\ref{fig-all}(a)), which we ascribe to the establishment of the long-range magnetic order.
The integrated intensity of the most pronounced magnetic reflection, i.e., at $Q$\,=\,1.712\,\AA$^{-1}$, exhibits a clear order-parameter-like response (Fig.~\ref{fig-all}(b)), i.e., $\propto$ $(T-T_N)^{2p}$, where $T_N$\,=\,54.0(2)\,K is the magnetic transition temperature and $p$\, = 0.22(2) is the critical exponent. This value of the critical exponent is close to that expected for an \textit{XY} model \cite{PhysRevLett.92.177201}.
We point out that the factor of 2 in the exponent accounts for the fact that the magnetic-reflection intensity is proportional to square of the magnetic moment.\\ \\
In order to determine the magnetic structure, we first subtracted a paramagnetic signal from the base temperature data, by subtracting the 60 K data from the 1.5 K data.
The derived difference plot is shown in Fig.~\ref{fig-all}(c).
The position of the magnetic reflections can be described with the (1~0~0) as well as (0~0~1) magnetic wave vectors. 
The representation analysis based on either of these two magnetic wave vectors gives the same four one-dimensional irreducible representations (IRRs), each consisting of three basis vectors that connect four crystallographically equivalent Co atoms within the unit cell (Table\,\ref{tab-irreps}), while the magnetic unit cell corresponds to the $C$\,-$1$ space group.
These basis vectors were then used as an input for the magnetic structure refinement to the difference between the 1.5\,K and 60\,K diffraction patterns.
From the four possible irreducible representations only $\Gamma_3$ agrees with the experiment, yielding a very high quality of the refinement (Fig.~\ref{fig-all}(c)) with $\psi_3^1$\,=\,$-2.07(3)$, $\psi_3^2$\,=\,$-0.32(5)$, and $\psi_3^3$\,=\,2.23(3).
The derived magnetic moment for Co at the atom\,1 site is thus ($m_a$,\,$m_b$,\,$m_c$)\,=\,$[-2.07(3),\,-0.32(5),\,2.23(3)]$\,$\mu_B$, where $\mu_B$ stands for Bohr magneton, yielding the ordered magnetic moment size of 3.3(1)\,$\mu_B$, i.e., in reasonable agreement with the Co$^{2+}$ $S$\,=\,3/2.
The corresponding magnetic structure, plotted in the inset in Fig.~\ref{fig-all}(b), indicates an almost collinear antiferromagnetic order aligned along $\sim$(1\,0,$-$1) with an additional small $b$ component alternating between the edge sharing dimers. 
 \begin{figure*}
	\centering
	\includegraphics[width=18 cm, height= 6 cm]{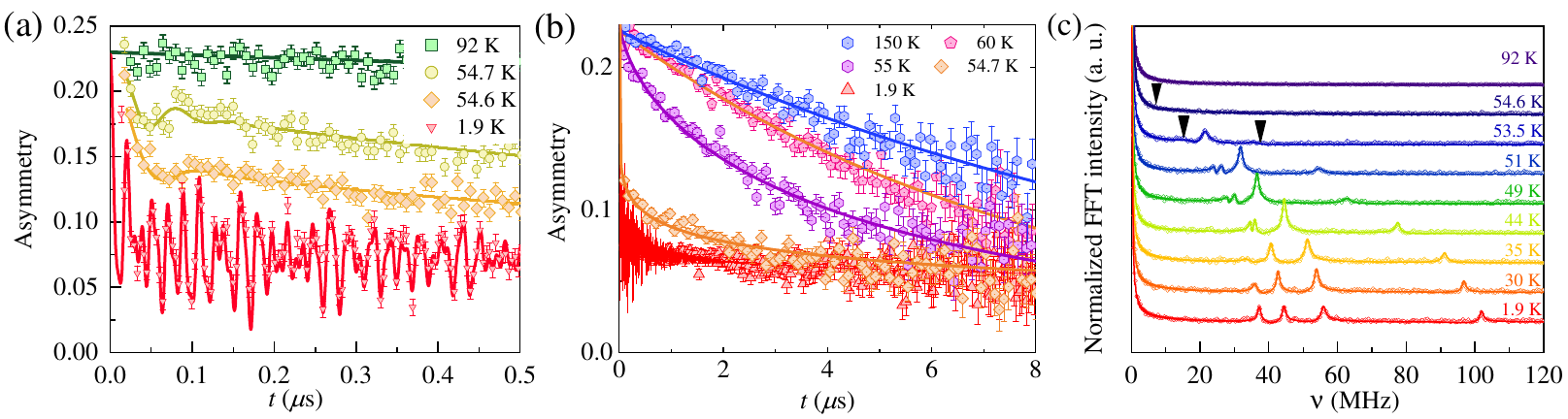}
	\caption{(a) Time evolution of muon spin asymmetry at short times at several representative temperatures in zero-field. The solid red lines represent fits to the function described in the text. (b) Time evolution of muon spin asymmetry at longer times in zero-field.  (c) Fourier transform of the time-dependent asymmetry depicted in (a). The arrow indicates the position of oscillating frequencies corresponding to the respective temperatures. }{\label{osMUSR}}.
\end{figure*}
\begin{figure}
	\centering
	\includegraphics[width=0.5\textwidth]{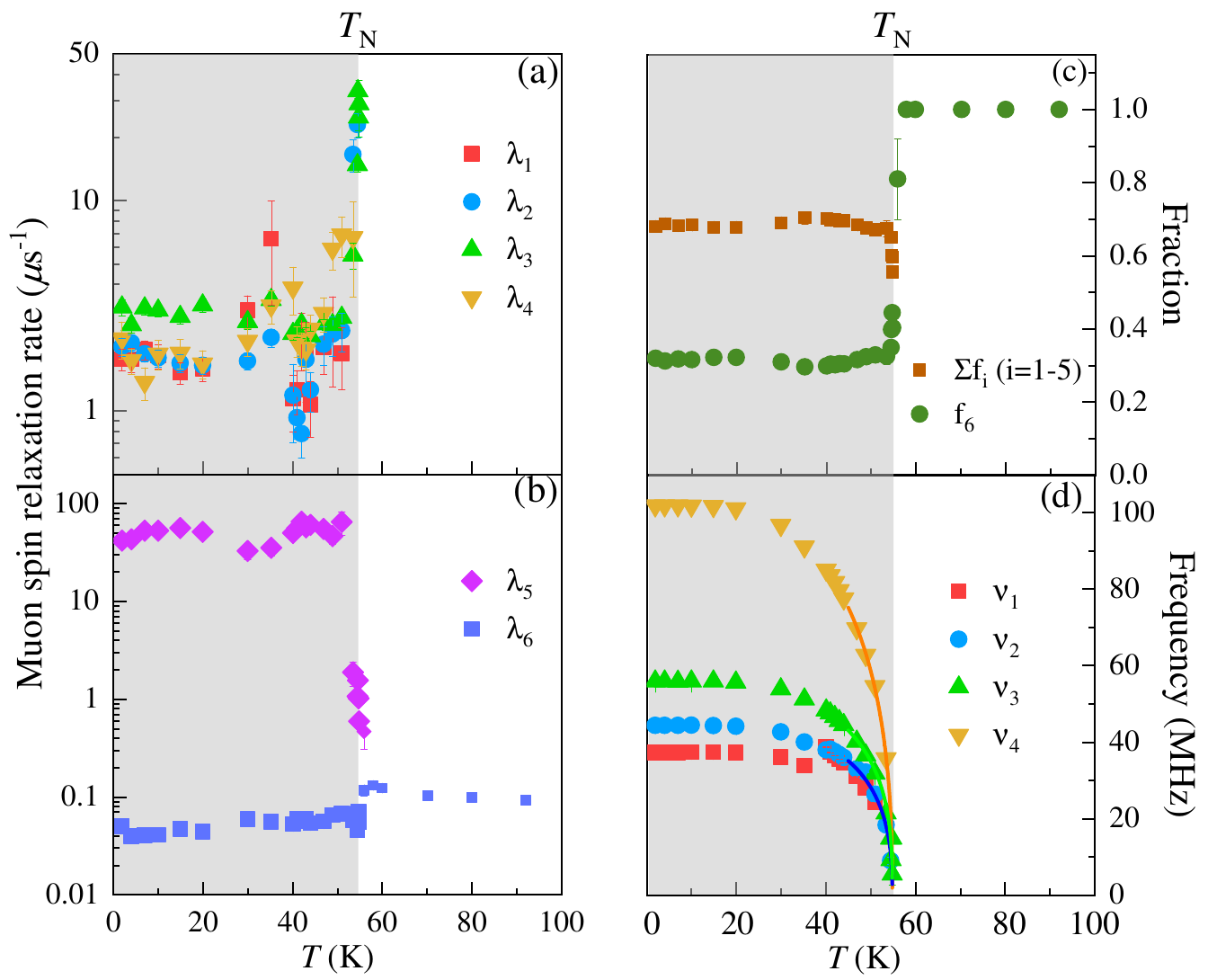}
	\caption{(a) Temperature dependence of  muon spin relaxation rates of transverse components below $T_{\rm N}$ within the short time scale in zero-fields. (b) Temperature dependence of transverse and longitudinal muon spin relaxation rates over a longer time scale. (c) Temperature dependence of the sum of the factions of transverse and longitudinal components in zero-field. (d) Temperature dependence of four muon frequencies. The solid  line represents the  critical scaling behavior of muon frequency.  }{\label{CTOFig6}}.
\end{figure}
\subsection{Muon spin relaxation}
In order to get further insights into the spin dynamics and local order parameter in the ground state of CTO, $\mu$SR measurements were performed at several temperatures in zero-field (ZF). Figure~\ref{osMUSR}(a) shows the time evolution of the $\mu$SR spectra at a few selected temperatures at short times. Upon lowering the temperature below $T_{\rm N}$, the appearance of clear coherent oscillations at low temperatures within the early $\mu$SR time window indicates the presence of static internal fields, consistent with the onset of a long-range magnetically ordered state in this compound.\\ \\ For polycrystalline materials exhibiting long-range magnetic order, when the internal magnetic field at the muon site ($B_{\rm loc}$) is uniform, the time evolution of the muon spin asymmetry, $P(t)$, typically follows a characteristic form
$P(t) = \frac{1}{3} + \frac{2}{3}\cos(\omega_{\mu}t)$ below $T_{\rm N}$.
In this expression, the oscillating term with frequency $\omega_{\mu}$ reflects the precession of the muon spins around the local magnetic field. The frequency $\omega_{\mu}$ is directly related to the magnitude of the internal field through the relation $\omega_{\mu} = \gamma_{\mu} B_{\rm loc}$, where $\gamma_{\mu} = 2\pi \times 135.5$~MHz/T is the muon gyromagnetic ratio. The oscillating (cosine) component, associated with the prefactor $2/3$, arises from the muon spin components that are initially perpendicular (transverse) to the local internal field. While, the non-oscillating $1/3$ term corresponds to the muon spin components that are aligned parallel (longitudinal) to $B_{\rm loc}$ and thus do not experience precession. This $1/3$--$2/3$ division arises because, in a polycrystalline sample, the local internal fields at the muon stopping sites (typically located about 0.9–1.1~Å away from an oxygen ion) are randomly oriented relative to the initial muon spin polarization.
\\ \\
In the present system, spin fluctuations cause the muon spin polarization to lose perfect coherence, leading to relaxation of both the longitudinal and transverse components. Above $T_{\rm N}$, rapid spin fluctuations in the motional narrowing regime effectively average out the distinct $1/3$ and $2/3$ contributions, resulting in a simple single-exponential relaxation without coherent oscillations  (Fig.~\ref{osMUSR}(b)). Upon lowering the temperature below $T_{\rm N}$, the muon spin polarization deviates from single-exponential behavior and instead exhibits damped oscillations, indicating the onset of a long-range magnetically ordered state. The presence of an oscillating signal over a short time scale (Fig.~\ref{osMUSR}(a)), combined with a non-oscillating 1/3 time dependent  signal over a longer time scale (Fig.~\ref{osMUSR}(b)), in the present compound suggests the coexistence of both static and fluctuating local magnetic moments below $T_{\rm N}$. \\ \\ To confirm the origin of damped oscillating signal, the temperature dependence of Fourier-transformed $\mu$SR spectra was obtained and is shown in Fig.~\ref{osMUSR}(b).  At the lowest measured temperature of $T = 1.9$~K ($\ll T_{\rm N}$), 
the Fourier-transformed $\mu$SR spectrum reveals four distinct peaks centered around 90~MHz, 55~MHz, 45~MHz, and 35~MHz (Fig.~\ref{osMUSR}(c)). 
The presence of multiple well-defined frequencies indicates the existence of several distinct oscillating components. 
These signals likely arise from muons experiencing different static local
magnetic fields within the sample, corresponding to internal fields of approximately 2~mT, 3~mT, 6~mT, and 8~mT, respectively. Such a distribution of fields suggests a complex magnetic ground state, potentially involving multiple magnetically inequivalent muon stopping sites 
or a multi-$\mathbf{q}$ magnetic structure. 
Our neutron diffraction results reveal a single-$\mathbf{q}$ magnetic structure, characterized by a propagation vector of the (100) or (001) type, indicating that the observation of four distinct muon precession frequencies does not originate from a multi-$\mathbf{q}$ scenario. Furthermore, multiple muon precession frequencies have been reported in several ordered magnets, even for a single muon stopping site, due to the breaking of the threefold symmetry around magnetic ions surrounded by oxygen ions~\cite{PhysRevB.83.024414,McClarty2020}. In addition, in Na$_{2}$Co$_{2}$TeO$_{6}$, two distinct muon oscillation frequencies have been attributed to the presence of a multi-domain magnetic structure \cite{Jiao2024}. In the present system, the observation of four distinct muon precession frequencies can be associated with the complex magnetic structure as revealed by neutron diffraction. The canted antiferromagnetic alignment, along with the slight alternation of the $b$-axis component between edge-sharing dimers, may lead to multiple inequivalent magnetic environments within the unit cell. Consequently, muons occupying similar crystallographic positions could experience slightly different static internal fields, giving rise to the observed frequencies. However, a more definitive understanding of the origin of the four frequencies would require DFT calculations to identify the most stable muon stopping sites and their corresponding local magnetic fields, which besides being very demanding computationally, such task is beyond the scope of the present manuscript.  \\ \\ Upon increasing the temperature toward $T_{\rm N}$, the precession frequencies gradually shift to lower values. For instance, below $T = 53.5$~K, four distinct oscillating components are observed. As the temperature approaches $T_{\rm N} = 55$~K, specifically in the range $54.6\,\text{K} < T < T_{\rm N}$, only a single oscillating component remains, as shown in Fig.~\ref{osMUSR}(c). 
This progressive shift of the precession frequencies toward lower values as $T_{\rm N}$ is approached reflects the gradual weakening of the static internal magnetic fields, associated with the thermal suppression of the ordered moment size, similar to the behavior expected for an order parameter near a magnetic phase transition. \\
\\ In order to find the temperature variation of 
 relaxation rate and frequency of muon oscillations, the zero-field $\mu$SR data (solid line in \ref{osMUSR}(a) and (b)) were modeled using the polarization function  
 \begin{align}\label{Mufit}
 P_{\rm CTO}(t) &= \sum_{i = 1}^{4} f_{i} \cos(2\pi\nu_{i}t+\pi\phi/180) \exp(-\lambda_{i}t) \notag \\
 &\quad + f_{5} \exp(-\lambda_{5}t) + \textit{f}_{6} \exp(-\lambda_{6}t),
 \end{align}
 over the entire measured time window.  
 This relaxation function incorporates both the transverse ($2/3$) and longitudinal ($1/3$) components, accounting for both the short- and long-time behaviors.  
 The first term describes the four oscillating transverse components, characterized by their respective fractions $f_{i} (i = 1$-$4)$, frequencies $\nu_{i} (i = 1$-$4)$, and transverse relaxation rates $\lambda_{i} (i = 1$-$4)$.  
 The second term represents an additional non-oscillating transverse component, with fraction $f_{5}$ and transverse relaxation rate $\lambda_{5}$, which becomes prominent at longer times.  
 The third term captures the longitudinal relaxation, associated with fraction $f_{6}$ and longitudinal relaxation rate $\lambda_{6}$, persisting over longer time scales. \\ \\
 The estimated temperature dependence of the four  transverse relaxation rates at the early times   is presented in Fig.~\ref{CTOFig6}(a). At low temperatures $T<< T_{\rm N}$, the relaxation rates $\lambda_{i}$ exhibit a finite and temperature-independent behavior, reflecting the presence of persistent spin dynamics even in the ordered state. However, as the temperature approaches $T_{\rm N}$, a sharp increase in $\lambda_{i}$ is observed, indicating an enhancement of thermal fluctuations near $T_{\rm N}$. A similar behavior has been observed in a several Kitaev materials below $T_{\rm N}$ including RuBr$_{3}$ \cite{PhysRevB.109.014440} and Na$_{2}$Co$_{2}$TeO$_{6}$ \cite{Jiao2024}. \\ \\ The temperature dependence of the muon spin relaxation rates $\lambda_{5}$ and $\lambda_{6}$, associated with the transverse and longitudinal components dominating the longer time scales, is shown in Fig.~\ref{CTOFig6}(b). Below $T_{\rm N}$, both $\lambda_{5}$ and $\lambda_{6}$ remain nearly constant, with $\lambda_{5} \gg \lambda_{6}$. The nonzero, temperature-independent relaxation rates below $T_{\rm N}$ indicate persistent dynamic local fields, a feature commonly observed in frustrated magnetic systems. As the temperature approaches $T_{\rm N}$, a sharp increase  in $\lambda_{5}$ is observed, suggesting the onset of critical slowing down near the magnetic transition. Above $T_{\rm N}$, the muon spin polarization follows a single-exponential form with a fraction close to one and a constant relaxation rate $\lambda_{6}$, consistent with the fast fluctuating (motional narrowing) regime. \\ \\
Figure~\ref{CTOFig6}(c) shows the temperature dependence of the fractions associated with the muon spin polarization components. This includes four oscillating signals with fractions \(f_1, f_2, f_3,\) and \(f_4\) at short time scales, as well as non-oscillating components with fractions \(f_5\) and \(f_6\) at longer time scales. The sum of fractions from \(f_1\) to \(f_5\) (denoted as \(\Sigma f_i\) for \(i = 1\) to \(5\)) is approximately 2/3, indicating these components originate from transverse relaxation processes in zero-field. Meanwhile, \(f_6\), which remains around 1/3, represents the longitudinal component.  
Below 54~K (shaded region), the fractions $\sum f_i$ and $f_6$ remain stable, indicating the coexistence of transverse and longitudinal components associated with static and dynamic local fields, respectively. As the temperature rises above 54~K, the system evolves into a cooperative paramagnetic state, where only the longitudinal component $f_6$ remains with a fraction close to one. This cooperative paramagnetic behavior, observed well above $T_{\rm N}$, is characteristic of classical magnetic systems and is further supported by thermodynamic results.
\\ \\
\begin{table}[t]
	\centering
	\caption{The estimated empirical parameters based on the order-parameter-like behavior of the muon precession frequency for the present compound, along with literature values for other compounds, are summarized.}
	\begin{tabular}{|c|c|c|c|c|c|}
		\hline \hline
		Compound & $T_{\rm N}$ &	$\nu(0)$ & $H_{\rm loc}(0)$ & $\alpha$ & $\beta$ \\ 
		& (K) & MHz & (mT) & & \\ \hline
		Co$_{2}$Te$_{3}$O$_{8}$ & 54.8 & 102.13(15)  & 753.7  & 3.27(5) & 0.41(0) \\
		($C2/c$) &  & 56.24(69) & 415.05 & 3.08(35) & 0.33(1) \\ 
		Present & &44.24(31) & 326.49 & 4.01(30)  & 0.38(1) \\ \hline
		RuBr$_{3}$ & 34.19 & 4.61 & 34.08  & 2.45(19) & 0.32(1) \\ 
		($R\bar{3}$,\cite{PhysRevB.109.014440}) &  & 3.94 & 29.09 & 2.44(18) & 0.31(1) \\ \hline
		Na$_{2}$Co$_{2}$TeO$_{8}$ & 25.8 & 3.85(2)  & 13 & 1.93 & 0.47(5) \\ 
		($P21$,\cite{Jiao2024}) & & 3.85(2)  & 12.7 & 3.16 & 0.47 \\ \hline
	\end{tabular}\label{expone}
\end{table}
Figure~\ref{CTOFig6}(d) depicts the temperature dependence of muon frequencies below $T_{\rm N}$. The shifting of frequencies to lower values indicates that the effective internal fields weaken as the transition temperature \(T_{N}\) is approached, likely due to critical fluctuations and the gradual alignment of magnetic moments into a more ordered configuration. It is worth noting that the gradual merging of $\nu_1$ and $\nu_2$ around 40~K may suggest the onset of a slow spin reorientation process, where the spin configuration evolves with temperature, leading to increasingly similar local magnetic environments sensed by the muons. This behavior results in the progressive convergence of the two precession frequencies. However, no clear evidence for a spin reorientation is observed in our neutron diffraction data, indicating that if such a reorientation occurs, it likely involves subtle changes that are not easily detected by diffraction techniques. Further detailed studies would be necessary to confirm this possibility. The observed frequencies, which reflect the local internal magnetic fields, were modeled using the phenomenological expression
\begin{equation}\label{order}
\nu_{i}(T) = \nu_{i}(0) \left[1 - \left(\frac{T}{T_{N}}\right)^{\alpha}\right]^{\beta},
\end{equation}
where $\nu_{i}(0)$ is the muon precession frequency at zero temperature, $\alpha$ is an empirical parameter controlling the saturation behavior, and $\beta$ is the critical exponent like parameter. The solid lines in Fig.~\ref{CTOFig6}(d) show that the temperature dependence of the muon frequencies follows an order-parameter-like behavior according to Eq.~\ref{order} below $T_{\rm N}$. The fitted values of $\alpha$, $\beta$, and $\nu_{i}(0)$ for a fixed $T_{\rm N}$ are summarized in the table~\ref{expone}. 
The $\beta$ exponent obtained from $\mu$SR represents a phenomenological fitting parameter rather than a direct measure of the magnetic order parameter. This accounts for the discrepancy with the critical exponent extracted from neutron diffraction, which directly reflects the order parameter.

\begin{figure*}
	\includegraphics[width=\textwidth]{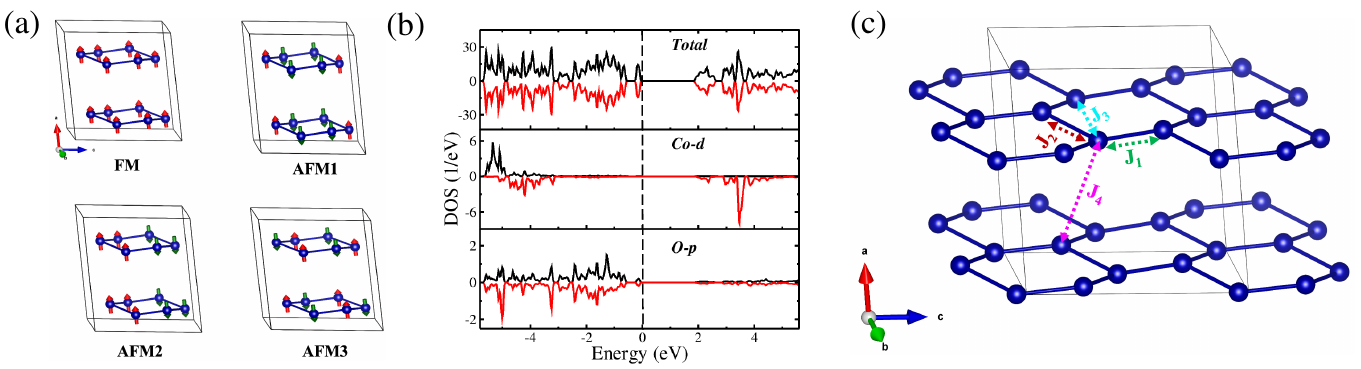}
	\caption{(a) Schematic illustration of four magnetic configurations between nearest-neighbor Co ions in Co$_2$Te$_3$O$_8$: ferromagnetic (FM) ordering, antiferromagnetic (AFM1), AFM2, and AFM3. These configurations were considered to identify the lowest energy magnetic ground state.
		 (b) Computed spin-polarized total and partial density of states of Co-\(d\) and O-\(p\) states within the lowest-energy configuration. Here the Fermi energy labeled as a 0~ev. (c) Schematic depicts nearest-neighbor intra-planer and inter-planer exchange interactions.}
	\label{mag_confi}
\end{figure*}
\begin{table}[t]
	\centering
	\centering
	\caption{Total energies and magnetic moments of various magnetic configurations of Co$_2$Te$_3$O$_8$ obtained using GGA and GGA+\(U\) calculations. Energies are given relative to the FM configuration in units of meV per formula unit (F.U.), and magnetic moments (MM) are in units of \(\mu_\text{B}\) per Co atom.}
	\label{Lowest_energy}
	\begin{tabular}{lcc|cc}
		\hline\hline
		Config. & \multicolumn{2}{c|}{GGA} & \multicolumn{2}{c}{GGA+\(U\)} \\
		& Energy & Moment & Energy & Moment \\
		& (meV/F.U.) & (\(\mu_\text{B}\)) & (meV/F.U.) & (\(\mu_\text{B}\)) \\
		\hline
		FM    & 0.00   & 2.57 &  0.00   & 2.75 \\
		AFM1  & -30.00 & 2.53 & -12.58  & 2.75 \\
		AFM2  & -27.50 & 2.55 & -8.38   & 2.75 \\
		AFM3  & -67.50 & 2.51 & -21.82  & 2.75 \\
		\hline\hline
	\end{tabular}
\end{table}
\section{Calculated exchange interactions}
We begin by examining the lowest energy magnetic state and electronic structure of bulk Co$_2$Te$_3$O$_8$.  Several magnetic configurations, as schematically shown in Fig.\ref{mag_confi}, were considered to identify the nature of magnetism in this system. The total energies corresponding to these spin arrangements are summarized in Table\ref{Lowest_energy}. Our calculations indicate that an antiferromagnetic configuration AFM3 is energetically most favorable within both GGA and GGA+U approaches, in agreement  with magnetic structure determined by neutron diffraction as well as with experimental observations of negative CW temperature. The computed energy differences $\Delta E = E_\text{FM} - E_\text{AFM}$ are significantly larger in GGA compared to GGA+$U$, reflecting the known tendency of the latter to reduce strength of inter-atomic magnetic exchange couplings due to enhanced localization of Co $3d$ electrons~\cite{anisimov1997first} as also observed in other transition metal oxides. For instance, the energy difference between FM and AFM3 configurations is reduced from 67.5 meV/F.U. in GGA to 21.8 meV/F.U. in GGA+$U$. Thus, our results demonstrate that antiferromagnetic Co-Co couplings are robust with respect to the choice of theoretical approaches. However, it is important to incorporate Hubbard $U$ in our calculations since $3d$ orbitals are highly localized and GGA is known to underestimate the insulating gap of such oxides. The rest of the analysis is presented from our GGA+U calculations. 
\\ \\
To gain further insight into the nature of the ground state electronic structure, we analyzed the total and atom-projected density of states (DOS), shown in Fig.~\ref{mag_confi}(b), for the AFM3 configuration. The total DOS shows that the system is an insulator with a band gap of approximately 1.9 eV. The partial DOS of Co-$3d$ states displays a clear exchange splitting between spin channels. The majority spin states are fully occupied, while the minority spin states are largely filled, pointing to a high-spin $d^7$ configuration for Co$^{2+}$ ions, consistent with thermodynamic results. This is further supported by the calculated spin moment of 2.75 $\mu_{\rm B}$ at each Co site. Additionally, the O-$2p$ states are spread throughout the valence band region, signifying substantial Co–O hybridization. This hybridization is responsible for the slight deviation of spin-moment of Co from its ionic value of 3.0 $\mu_B$, arising from the three unpaired electrons of Co-3$d$ orbitals. Thus both strong electron correlation and covalency play crucial roles in achieving the insulating ground state of  this system. Since O-$p$ states arises just below the Fermi level, this system belongs to a charge transfer insulator (see Fig.\ref{mag_confi}(b)). \begin{table}[b]
	\caption{Calculated inter-atomic exchange interactions \(J_{ij}\) at ambient conditions using GGA+$U$. Here, NN denotes the number of equivalent nearest-neighbor pairs for each interaction.}
	\label{table-1}
	\centering
	\renewcommand{\arraystretch}{1.4}
	\setlength{\tabcolsep}{12pt}
	\begin{tabular}{cccc}
		\toprule
		\(\mathbf{J_{ij}}\) & \textbf{NN} & \textbf{Distance (Å)} & \textbf{Energy (meV)} \\
		\midrule
		\(J_1\) & 1 & 3.47 & -1.72 \\
		\(J_2\) & 2 & 3.57 & -3.07 \\
		\(J_3\) & 2 & 5.21 & -0.34 \\
		\(J_4\) & 1 & 5.96 & -0.07 \\
		\bottomrule
		\hline\hline
	\end{tabular}
\end{table}
\par 
Further to understand the origin of the experimentally observed antiferromagnetic ordering in Co$_2$Te$_3$O$_8$, we calculated the interatomic exchange interactions using the magnetic force theorem developed by Liechtenstein et al. \cite{PhysRevB.61.8906}.
Our analysis identifies four significant nearest-neighbor exchange pathways, labeled $J_1$ through $J_4$ (marked in Fig.~\ref{mag_confi}(c)), with distinct interaction strengths and interatomic separations (see Table~\ref{table-1}). The leading couplings, $J_1=-1.72$ meV and $J_2=-3.07$ meV, are both antiferromagnetic and relatively strong, involving Co--Co distances of 3.45~\AA{} and 3.57~\AA{}, respectively. These couplings likely drive the robust AFM alignment between neighboring Co spins, consistent with the experimentally observed long-range antiferromagnetic order. In contrast, the more distant interactions, $J_3=-0.21$ meV and $J_4=-0.05$ meV, are significantly weaker but still retain an antiferromagnetic character, suggesting that frustration and long-range coupling effects are marginal. The mean-field estimate of the Néel temperature is 66~K, in good agreement with experiment. This hierarchy of exchange interactions suggests that the magnetism in Co$_2$Te$_3$O$_8$ can be described predominantly by a Heisenberg model with AFM couplings. Notably, the structural motif of edge-sharing CoO$_6$ octahedra facilitates these superexchange interactions via oxygen-mediated paths. Taken together, our calculated exchange parameters not only corroborate the experimentally determined AFM ground state but also provide a microscopic basis for the observed magnetic ordering. These insights are essential for developing effective spin models and can guide future studies, including spin dynamics simulations or inelastic neutron scattering investigations of the magnon spectrum.
\section{CONCLUSION}
In summary, we have investigated the ground state properties and spin dynamics of polycrystalline Co$_{2}$Te$_{3}$O$_{8}$, where Co$^{2+}$ ions form a two-dimensional honeycomb lattice stacked along the $a$-axis, using a combination of thermodynamic measurements, neutron diffraction, $\mu$SR spectroscopy, and DFT calculations. Our combined thermodynamic and microscopic results, well supported by DFT calculations, reveal the presence of long-range antiferromagnetic order below $T_{\rm N} \approx 55$~K. Below $T_{\rm N}$, the magnetic excitations are indicated by a $C_{\rm mag} \propto T^3$ behavior, characteristic of three-dimensional antiferromagnets with magnon excitations. $\mu$SR experiments uncover the local order parameter-like behavior of internal magnetic fields, which is consistent with neutron diffraction results and supports a \textit{XY}-type antiferromagnetic state. In addition, $\mu$SR measurements provide concrete evidence for the coexistence of static and dynamic local magnetic fields in the ground state, revealing persistent spin dynamics below $T_{\rm N}$ and critical slowing down of spin fluctuations near the transition. DFT calculations confirm that the long-range magnetic order is stabilized by intra-planar further-neighbor interactions and weak inter-planar antiferromagnetic couplings, highlighting the role of moderate spin frustration. Future  studies on the single crystals of Co$_{2}$Te$_{3}$O$_{8}$  may reveal critical behavior of spins in the pursuit of a  classical spin liquid state.

    \section*{Acknowledgments}
     P.K. acknowledges the funding by the Science and
    Engineering Research Board, and Department of Science and Te chnology, India through Research Grants. J.K. thanks M. Barik for her support in synthesizing a batch of the Co$_{2}$Te$_{3}$O$_{8}$ sample. The work at CALDES was supported by the Institute for Basic
    Science (IBS-R014-Y2). The work at SKKU was supported by the National Research Foundation
    (NRF) of Korea (Grant no. RS-2023-00209121, 2020R1A5A1016518). This work has been funded by the Slovenian Research Agency (project J2-2513, and program No. P1-0125). S.K.P acknowledges support from ANRF (Previous SERB), Government of India for the core research grant
    (CRG/2023/003063). Experiments at the ISIS Neutron and Muon Source were supported by beamtime allocation RB2310236 from the Science and Technology Facilities Council.
    \section{Data availability}
    The data that support the findings of the current study
    are available from the corresponding author upon reason
    able request.
\bibliography{CTO}
\end{document}